\documentclass[aps,prl,preprint]{revtex4}
\usepackage{amssymb}
\usepackage{bm}
\begin{document}
\noindent{\bf Comment on ``Highly Extended Image States around
Nanotubes"}

In their Letter \cite{granger}, Granger {\it et.al.} have
discussed the image states around carbon nanotubes. To do that,
they derive the interaction potential $V(\rho)=\frac{2q^2}{\pi
a}\sum\limits_{n=1,3,5,\cdots}li[(\frac{a}{\rho})^n]$ between a
nanotube of radius $a$ and an electron at a distance $\rho$ from
its axis. Then, they formally introduce the effective interaction
$V_{eff}(\rho)=V(\rho)+\frac{l^2-\frac{1}{4}}{2\mu \rho}$ and give
the  transverse Schr\"{o}dinger equation
\begin{equation}\label{trans}\{\frac{d^2}{d\rho^2}+2\mu[E_{nl}-V_{eff}(\rho)\}\psi_{nl}(\rho)=0.\end{equation}

In our viewpoint, the interaction potential $U$ between a nanotube
and an electron must be the function of $\rho$, $\phi$ and $z$ in
terms of the lattice structure of the carbon nanotube. Here
$\rho$, $\phi$ and $z$ are the components of cylindrical
coordinate system whose $z$-axis is the tube axis. Besides, $U$
satisfies $U(\rho,\phi+2\pi j_1/N,z+j_1\tau+j_2T)=U(\rho,\phi,z)$
because the nanotube's symmetry\cite{saito,tzc}. Here $j_1, j_2$
are integer numbers, and $N, \tau, T$ are characteristic
quantities of the nanotube, e.g., for (10,10) nanotube, $N=20,
\tau=1.23\text{\AA}, T=2\tau$. When $\rho\gg a$, it seems rational
to approximate $U(\rho,\phi,z)\approx V(\rho)$. But, there is no
reason to think $U(\rho,\phi,z)\approx V(\rho)$ when $\rho\sim a$.
Therefore, we believe that the energies $E_{nl}$ obtained from
Eq.(\ref{trans}) are merely valid for $n,l\gg 1$.

If we know $U(\rho,\phi,z)$, we can calculate the energies of
image states in principle. From the periodical structure of $U$,
we can obtain the general Bloch's theorem\cite{tzc}
\begin{equation}\label{bloch}\begin{array}{l}H\varphi(\rho,\phi,z)=E\varphi(\rho,\phi,z),\\
\varphi(\rho,\phi+\frac{2\pi
j_1}{N},z+j_1\tau+j_2T)=e^{i[\frac{2\pi lj_1}{N}+\kappa
(j_1\tau+j_2T)]}\varphi(\rho,\phi,z),\end{array}\end{equation}
where $l=0,1,\cdots,N-1$ because of the periodical boundary
condition along circumference and $0\leq k<2\pi/T$. Adopting the
thought of plane wave expansions \cite{callaway},
$\varphi_{nl\kappa}(\rho,\phi,z)=\sum\limits_{j_1j_2}C_{j_1j_2}\chi_{nl\kappa}(\rho)e^{i[(l+j_1N-j_2N\tau/T)\phi+(\kappa+2\pi
j_2/T)z]}$, we substitute it in
$H\varphi_{nl\kappa}(\rho,\phi,z)=E_{nl\kappa}\varphi(\rho,\phi,z)$
and obtain the transverse Schr\"{o}dinger equation
\begin{equation}\label{our}det[(\mathcal{H}_{m_1m_2j_1j_2}-E_{nl\kappa}\delta_{j_1m_1}\delta_{j_2m_2})\chi_{nl\kappa}(\rho)]=0,\end{equation}
where
$\mathcal{H}_{j_1j_2m_1m_2}=\mathcal{T}_{j_1j_2}\delta_{j_1m_1}\delta_{j_2m_2}+\mathcal{U}_{j_1j_2m_1m_2}$,
$\mathcal{T}_{j_1j_2}=-\frac{\hbar^2}{2\mu}[\frac{d^2}{d\rho^2}+\frac{d}{\rho
d\rho}-\frac{(l+j_1N-\frac{j_2N\tau}{T})^2}{\rho^2}-(\kappa+\frac{2\pi
j_2}{T})^2]$, and $\mathcal{U}_{j_1j_2m_1m_2}=\frac{N}{2\pi
T}\int_0^{2\pi}d\phi\int_0^TdzU(\rho,\phi,z)e^{i[(j_1-m_1)-\frac{(j_2-m_2)\tau}{T}]N\phi+\frac{i2\pi
(j_2-m_2)z}{T}}$.

Hence, the key problem is to determine the potential
$U(\rho,\phi,z)$. It is an extremely difficult project beyond this
comment.

\vspace{1cm}
\noindent Z. C. Tu$^1$ and Z. C. Ou-Yang$^{1,2}$\\
\indent$^1$Institute of Theoretical Physics\\
\indent The Chinese Academy of Sciences\\
\indent P.O.Box 2735 Beijing 100080, China\\
\indent $^2$Center for Advanced Study\\
\indent Tsinghua University, Beijing 100084, China\\

\noindent PACS numbers: 61.46.+w, 34.60.+z, 34.80.Lx, 36.10.¨Ck

\end{document}